\documentclass[reqno,11pt]{article}
\usepackage[]{graphicx}
\usepackage{amsfonts,amssymb,amsmath,amscd,cite,palatino}

\def\be{\begin{equation}}
\def\beq{\begin{equation}}
\def\ee{\end{equation}}
\def\eeq{\end{equation}}
\def\ba{\begin{eqnarray}}
\def\ea{\end{eqnarray}}
\def\eqn#1{\begin{equation}\begin{split}#1\end{split}\end{equation}}

\def\P{\partial}

\def\t{\theta}

\def\text#1{{\hbox{#1}}}
\def\1{{-1}}
\def\2{\frac{1}{2}}
\def\<{\langle}
\def\>{\rangle}

\def\d{\delta}

\def\IP{\relax{\rm I\kern-.18em P}}

\begin{document}

\vskip.8cm
\vskip.7cm
\begin{Large}
\centerline{{\bf Backgrounds in Boundary String Field Theory} }  
\end{Large}
\vskip.2cm
\centerline{by}
\vskip.2cm
\centerline{{\bf 
Marco Baumgartl\footnote{E-mail: \sf M.Baumgartl@physik.uni-muenchen.de}}
}
\vskip 5mm
\centerline{\it Arnold-Sommerfeld Center for Theoretical Physics (ASC)}
\centerline{\it Ludwig-Maximilians-Universit\"at M\"unchen, Theresienstr. 33, D-80333 Munich, Germany}
\vskip 5mm
\centerline{\it Cluster of Excellence for Fundamental Physics ``Origin and Structure of the Universe''}
\centerline{\it Boltzmannstr. 2, D-85748 Garching, Germany}
\vskip.7cm
\vskip1.2cm
\centerline{{  \bf Abstract} }
\par\noindent
\vskip.8cm
We study the role of closed string backgrounds in boundary string field theory. Background independence requires the introduction of dual boundary fields, which are reminiscent of the doubled field formalism. We find a correspondence between closed string backgrounds and collective excitations of open strings described by vertex operators involving dual fields. Renormalization group flow, solutions and stability are discussed in an example. 
\par\noindent
Contribution to proceedings of SFT09 in {\it Theoretical and Mathematical Physics}, Russian Academy of Sciences

\par\noindent

\vfill\eject

\section{Introduction}

Solutions of string field theory are usually given by two-dimensional conformal field theories, describing the string worldsheet embedded in some target space. At criticality string excitations are described by operators in the corresponding conformal field theory. Away from the critical point, not too much is known about string theory. In particular, each target space background which admits a conformal field theory should be a solution of a string field theory action. In general, however, it is unclear how different target space geometries are related in their worldsheet descriptions.

The presence of open strings requires worldsheets with boundaries, which have disk topology in lowest order tree-level. The boundary conditions which must be imposed often have a geometrical interpretation as D-branes embedded into the target space. They come with a variety of moduli, encoding their size and shape. It seems obvious that these moduli  and therefore the open string spectrum crucially depend on the closed string background. This suggests to view closed strings as the relevant objects that define a string theory, while open strings play a less fundamental role.

On the other hand 
there are reasons in favor for a fundamental role of open strings, too.
For example, the construction of a closed string field theory is generally much more complicated as an open string field theory \cite{Zwiebach:1997fe}. In addition there are well known connections between scattering amplitudes of open and closed strings, and in some cases open-closed dualities are explicitly known. 
If open strings are really the more fundamental objects, this raises the question in what sense an open string field theory may be independent of closed string backgrounds.

The most important question here is how a closed string background manifests itself in an OSFT. In the BSFT approach \cite{Witten:1992qy,Witten:1992cr,Shatashvili:1993kk,Shatashvili:1993ps} generalized conformal field theories are used in order to construct a consistent open string field theory -- generalized in the sense that while the closed string background and therefore the bulk CFT is kept on-shell and fixed, the fields inserted at the boundary, though, are not restricted by the requirement of conformal invariance. 
This construction can be carried beyond the case of flat boson CFTs. Non-trivial closed string backgrounds can be implemented by replacing the flat boson by (products of) CFTs which represent other target space geometries. In \cite{Baumgartl:2004iy,Baumgartl:2006xb,Baumgartl:2008st,Baumgartl:2008js} this has been done for group-valued WZW models. With the construction presented there BSFTs in different backgrounds can be compared to each other and the impact of background shifts can be studied.

An important result that has been obtained in \cite{Baumgartl:2004iy} is a `factorization property' shared by the flat boson and the WZW CFTs. It has been proven that the generalized partition function appearing in the BSFT construction factorizes into a boundary part, which captures all (off-shell) open string operators, and the rest, which has no coupling to open strings.  This kind of factorization does not require an integration over bulk-boundary terms, so that it exhibits a fundamental independence of bulk and boundary terms. This independence will be important for the comparison of BSFTs. It motivates a modification in the definition of the string field action, which will accomplish the desired background independence.

After a short introduction to BSFT we will review the factorization property of the boundary $\sigma$-model partition function. This is the key ingredient for the main result, which is the representation of closed string backgrounds in terms of open string excitations. We show that shifts in the closed string background require the introduction of dual boundary fields,  which are reminiscent of the doubled field formalism \cite{Tseytlin:1990nb,Dabholkar:2005ve,Hull:2006va,Hull:2009mi}. We will discuss the renormalization of certain such operators in an example.

\section{Backgrounds in BSFT}

We start with a classical configuration of string theory, which allows a description as a two-dimensional quantum field theory with $\sigma$-model action and conformal invariance. The worldsheet has a boundary on which open string operators will be inserted. 
In the proposal, that has been put forward in \cite{Baumgartl:2004iy, Baumgartl:2006xb} one starts with the action
\eqn{
	I=\int_\Sigma L_p(G_{\mu\nu},B_{\mu\nu}; X^\mu)+\oint_{\partial\Sigma} B(t^I; X^\mu)\ . 
}
The first term $L_p$ defines the closed string background. It depends on closed string couplings like the gravitational field $G_{\mu\nu}$ and the Kalb-Ramon anti-symmetric field $B_{\mu\nu}$, but also on heavy modes. Conformal invariance gives constraints on these fields in the form of RG $\beta$-functions.

The boundary conditions which must be imposed on open strings, are encoded in the boundary integral $B(t^I)$. It formally depends on $t^I$, the set of all open string couplings. The associated operators are given by 
\eqn{
	V_I(X^\mu) = \frac{\partial B(t^I; X^\mu)}{\partial t^I}\Biggr|_{t^I=0} \ .
}
The boundary interaction term $B(t^I)$ contains tachyons $T(X)$, photons $A(X)_\mu $ and massive fields. Later we will argue that it is natural to include new types of fields as well, which do not have a direct interpretation as boundary fields of a BCFT.

In a string field theory it is necessary to give the theory a meaning when it is off-shell. 
In BSFT the prescription is to integrate over all possible boundary values of the $\sigma$-model fields $X(z,\bar z)$ while keeping the bulk conformal without any closed string vertex insertions. The  $\beta$-function equations for closed string couplings must be satiesfied, thus.
Since boundary perturbations will not affect the bulk conformal symmetry (at tree level), the boundary interaction term may indeed break conformal invariance without loosing the CFT description in the bulk.  

We quote the final result from \cite{Shatashvili:1993kk,Shatashvili:1993ps}. With assumptions on the decoupling of matter and ghosts, the BSFT action is given by
\eqn{
\label{BSFTdef}
	{\cal S}[t^I]&=(1-\beta^I\partial_I){\cal Z}[t^I], \
}
where $\beta_I$ are the $\beta$-functions associated to the couplings $t^I$.
The generating functional ${\cal Z}[t^I]$ of open string correlation functions is given by the path integral expression
\eqn{
	{\cal Z}[t^I] = \int D[X^\mu] e^{-I[X]}\ .
}

\subsection{Background independence}

The string field theory action ${\cal S}[t^I]$ is a function of the open string couplings $t^I$. Its solutions are given by conformal fixed points $t_*^I$. Implicitly, these solutions exhibit a closed string dependence, which enters through the choice of a background in form of a CFT.
Let us denote the configuration space of the background by ${\cal M}$, which includes fields like the graviton $G_{\mu\nu}$, Kalb-Ramon field $B_{\mu\nu}$ and dilaton $\Phi$.

A priori it seems that when we change the configuration from ${\cal M}$ to ${\cal M}'$ we must construct a new BSFT from scratch, depending on the new choice of the underlying CFT. From the viewpoint of boundary conformal field theories there may be $\beta$-functions for open-closed couplings, the spectrum of boundary fields may change completely when the closed string background is modified, D-brane solutions might become inconsistent or new D-branes might appear \cite{Fredenhagen:2006dn,Baumgartl:2007an,Gaberdiel:2008fn}.

It turns out that for BSFT this viewpoint is not entirely correct. In fact, unlike the boundary sector of CFTs, BSFT can be defined in a way so that it displays independence of the closed string background. 

Concretely, we start with a BSFT formulated on a closed string background ${\cal M}$ and would like to study the effect of transiting to a background ${\cal M}'$. Simple perturbation theory in the closed string moduli is not enough to obtain a meaningful result, because the form of the action usually depends on the choice of the fixed point. Rather it must be made sure that the generic structure of the appearance of closed string moduli in the open sector is global, or independent of the chosen starting point ${\cal M}$. 

For a large class of conformal backgrounds this is indeed possible. Key to this behaviour is the factorization property of WZW models proven in \cite{Baumgartl:2004iy}, which we will review briefly. 

The factorization property allows to split the generalized partition functions as
\eqn{
	{\cal Z}(s=s_*,t) = {\cal Z}_0(s=s*)\cdot \hat{\cal Z}(\hat s, t)
}
for $s=s_*$ a conformal fixed point. Here ${\cal Z}_0$ is a specific instanton partition function, which depends on closed string moduli $s$ only (its open string moduli are trivial). All the open string moduli $t$ appear only in $\hat {\cal Z}(\hat s, t)$, together with bulk induced moduli $\hat s$. These appear technically in a similar way as the pure open moduli $t$. If one forgets about their origin in the bulk CFT it is not entirely clear how to distinguish induced closed moduli from open moduli. Often $\hat s$ are couplings that involve certain `almost local' operators, which will be introduced below; but also this is not true in every case -- some fields like the constant Kalb-Ramon field are known to be equivalent to a constant photon field strength at the boundary. On the other hand there may also be closed string couplings which do not induce open string operators at the boundary, for example those coming from topological terms.

From (\ref{BSFTdef}) we see that the definition of the BSFT action is insensitive to the factor ${\cal Z}_0(s)$ in the generalized partition function. Thus it makes sense and is indeed reasonable to redefine the BSFT action by diving out ${\cal Z}_0(s)$. Because of the factorization property this is possible globally, and corresponds to a $t$-independent normalization at each point of the closed string moduli space. It represents a way to `covariantize' the BSFT action on the moduli space, since it tells us how to define a derivation with respect to closed string moduli $\hat s$ consistently.

\subsection{Factorization}

On the level of the path-integral, the factorization property can be derived by splitting the worldsheet fields in an appropriate way. For bosons on a torus this can be easily understood. A general string map $X^\mu(z,\bar z)$ can always be decomposed uniquely in the following way
\eqn{
\label{FBsplit}
	X^\mu(z,\bar z) = X_0^\mu(z,\bar z) + X_b^\mu(f)\ ,
}
where $X_0^\mu(z,\bar z)$ satisfied D0 boundary conditions and $X_b^\mu$ is a functional of the boundary data $f$ of $X^\mu$. When $f$ specifies the value of the string map restricted to the boundary, then $X_b^\mu(f)$ is given by the unambiguous extension of the function $f$ to the interior of the disk, governed by holomorphicity. To be concrete, we set
$X_0|_{\partial\Sigma} =0$ and assume that $f$ is given as a function of the boundary coordinate $\theta$ with the expansion $f(\theta)=\sum f_ne^{in\theta}$. The equations of motions $\Delta X_b^\mu=0$ then dictate the extension $X_b^\mu(z,\bar z) = f_0^\mu + \sum_{n>0}(f_nz^n + f_{-n} \bar z^n)$. When we split $f$ in a positively and negatively moded part $f^\pm$, the Polyakov action for the free boson takes the form
\eqn{
	I = I_0[X_0] + I[f] = \frac{R^2}{4\pi i\alpha'} \int \partial X_0\bar\partial X_0
		+ \frac{R^2}{4\pi i\alpha'} \int \partial f^+\bar\partial f^-\ .
}
Partial integration has been used to remove all terms that mix $X_0$ and $X_b$. Hence it is obvious that the path-integral satisfies
\eqn{
	{\cal Z} = \int D[X_0] e^{-I_0[X_0]} \cdot \int D[f] e^{-I[f]}\ .
} 

There are two effects which might spoil the factorization. First, one must be careful with the measure of the path-integral, since this might generally contribute by an anomaly term. For free bosons there is no such anomaly, but for example in WZW models such contributions play a crucial role \cite{Baumgartl:2006xb}.
The other possible problem arises from the question if the classical splitting holds also in the quantum theory. Again, for the flat boson this issue does not arise -- the non-trivial proof for WZW models though can be found in \cite{Baumgartl:2004iy}.

\subsection{Boundary action}

In the simplest example of bosons on a torus with radius $R$ the boundary interaction term can be expressed in terms of the boundary values $f$ of the string map as
\eqn{
\label{raddef}
	I[f] \propto R^2 \oint d\theta f^\mu(\theta)\partial_\theta \bar f^\nu(\theta) \delta_{\mu\nu}\ .
}
This defines the propagator at the boundary. Note that this expression is `almost local' in the following sense. Re-written in terms of the $X_b$, it takes the form
\eqn{
	\oint d\theta X_b^\mu(\theta) (H_{\mu\nu} X_b^\nu)(\theta)\ ,
}
where $H_{\mu\nu}$ denotes the Hilbert transform 
\eqn{
	H_{\mu\nu}(\theta,\theta') = \d_{\mu\nu}\frac{1}{4\pi i} \oint d\t' \sum_n e^{in(\theta-\theta')}|n|\ .
}
The effect of the Hilbert transformation is a flip of the relative sign between positive and negative modes, which came from right-movers and left-movers. This is very much reminiscent of the worldsheet realization of T-duality, therefore we denote the transformed `dual' boundary field often by
\eqn{
	*{X}^\mu \equiv (H X)^\mu\ .
}

In the case where the closed string background has flat geometry but is endowed with a Kalb-Ramond field $B_{\mu\nu}$ the factorization can be shown in the same way, as long as the field is constant. Explicitly, the boundary action in the presence of a $B$-field becomes
\eqn{
\label{bahb}
	I[f] \propto \oint d\theta X_b^\mu\partial_\theta *X_b^\nu \delta_{\mu\nu}
		+ \oint d\theta X_b^\mu\partial_\theta X_b^\nu B_{\mu\nu}\ .
}
The local fields at the boundary are given by $X_b[f]=f+\bar f$. Thus the constant Kalb-Ramond field induces a local term under the boundary integral. This reflects the fact that in flat non-compact space a constant $B$-field appears in the open string sector like a constant electromagnetic field strength. 
The first term mixes the local field with its dual and is proportional to the background metric. For non-constant metric and Kalb-Ramond field more such terms appear. A prototype of such theories is given by WZW models.

In WZW models the target space is a group manifold. While the factorization of path-integrals is obvious for bosons on a torus, it is a priori not clear if this works also in the WZW case. A natural ansatz for the splitting of the fields is
\eqn{
\label{WZWspl}
	g(z,\bar z) = g_0(z,\bar z)\, k(z, \bar z)\ ,
}
where $g_0$ obeys D0 boundary conditions $g_0|=1$ just like the field $X_0$ before; $k$ is a classical solution given by $k(z,\bar z)=h(z)\bar h(\bar z)$, a product of holomorphic fields, and should be compared to $X_b$. In the vicinity of 1 or, equivalently, at large radius (\ref{WZWspl}) becomes (\ref{FBsplit}). Just as in the free field case, the field $k$ depends only on the boundary data of the fields on the disk.

The WZW action does not split under this ansatz in two independent parts like the free boson action. Rather, there is a term present which mixes the two fields $g_0$ and $k$. Luckily, this term always has the form $J\,\P H(z)$, where $J$ is the WZW current $J=g_0\bar\P g_0$ and $H(z)$ is a holomorphic field depending on $k$ only. 

It turns out that the factorization property of the path-integral is not spoiled.
This is because any powers of $\int d^2 z J\,\P H(z)$ appearing in the path-integral do not contribute in correlation functions, even in the presence of a boundary. This result has been obtained in \cite{Baumgartl:2004iy},
where also the construction of an appropriate boundary action for the WZW case has been addressed. 

\subsection{Renormalization group flow and curved branes}

A central question, which addresses the issue of (closed string) background independence, is whether it is possible to capture genuine closed string effects in this way. For example a bosonic theory with flat target space always admits a spacefilling brane. In a curved target space this is not true. On a $S^3 \cong SU(2)$ target the maximally symmetric branes will be $S^2$ surfaces, and there is no spacefilling brane in this case. Of course it is always possible to take the large radius limit and view the $S^3$ theory as a deformation of the ${\mathbb R}^3$ theory\footnote{Issues of topology change in this example are briefly discussed in \cite{Baumgartl:2004iy}.}. Such an expansion around infinite radius allows to identify those boundary interaction terms $V_R = \frac{\P}{\P R} {\cal Z}$ which are responsible for curvature effects. 

To lowest order
\eqn{
	V_R &= \lambda \oint d\theta \epsilon_{abc} f^a \bar f^b \P_\theta f^c + {\cal O}(\lambda^2)\ ,
}
where $\lambda \sim \frac{1}{R}$ is the expansion parameter.
Interesting behavior is seen not before second order in $\lambda$. At this order a tachyon is excited and initiates a condensation process. 

Compared to similar condensation processes in flat space, the process changes qualitatively in the presence of $V_R$.
Instead of condensing to a flat lower dimensional brane, the corresponding RG equations are solved for {\it spherical branes}.
One finds that the RG flow excites tachyonic couplings of the form
\eqn{
	\oint \rho (X^2-c^2)^2\ ,
}
up to higher order terms. The $\beta$-functions of the couplings $\rho$ and $c$ are obtained as
\eqn{
	\beta_\rho &< 0 \\
	\beta_{c^2} &= 4(1-2\lambda^2c^2)\ .
}
The fact that $\beta_\rho<0$ ensures that the condensation process actually takes place, effectively creating a $\delta$-function in the path-integral localized on the brane geometry for infinite coupling. The new feature comes from non-trivial zeros of $\beta_{c^2}$, which are independent of $\rho$. The $\beta$-function vanishes for
\eqn{
	c^2 = \frac{1}{2\lambda^2}\ ,
}
which fixes the radius of the spherical brane in dependence of the new coupling $\lambda$.

The resulting spherical branes have shown to be stable in the sense that all tachyons, which could possibly condense further, can be set to zero consistently and are not excited anymore along the RG flow. The spherical branes obtained display a characteristic dependence on the coupling $\lambda$. The inverse of $\lambda$, however, is just the level of the WZW theory. The relation between the radius of the spherical brane and the WZW level therefore is exactly what is expected for branes which are described as conjugacy classes in $SU(2)$, as expected \cite{AlekseevMC}.

Although these results have been obtained in perturbation theory, they serve as strong test in favor of closed string background independence of BSFT.

\section{Outlook}

An interesting consequence of the construction presented above is that dual boundary fields $*X^\mu$ can be treated on the same footing as the usual local string fields. Both fields are not independent, though, and therefore it is not necessary to specify an integration measure in the path-integral separately for the dual fields. Without further modification of the quantization scheme we can use elementary methods and CFT technology to compute correlation functions.
The propagators in the enhanced field space $(X^\mu, *X^\mu)$ have been computed e.g.\ in \cite{Tseytlin:1990nb,Gerasimov:2001pg}
\eqn{
	\< X^\mu(\t) X^\nu(\t')\> &= \< *X^\mu(\t) *X^\nu(\t')\> = \delta^{\mu\nu} \sum_{k\ne 0} \frac{1}{|k|}e^{ik(\t-\t')} \\
	\< X^\mu(\t) *X^\nu(\t')\> &= \delta^{\mu\nu} \sum_{k\ne 0} \frac{1}{k}e^{ik(\t-\t')} = - \< *X^\nu(\t) X^\mu(\t') \>\ . \\
}
The existence of dual boundary fields is a qualitatively new feature and allows to enlargement of the space of couplings in the boundary theory. In particular, for any of the usual open string fields one can introduce a dual spacetime field. Their vertex operators will have the same properties (conformal dimensions, tensor structure) as their counterparts, therefore this will simply introduce a copy of the coupling space. 

More interesting, though, is the possibility of adding fields whose vertex operators depend on both, $X$ and $*X$ as generalization of (\ref{bahb}).  
For example it is possible to introduce a coupling ${\cal G}_{\mu\nu}$ that appears on the quadratic level as symmetric tensor
\eqn{
	\oint {\cal G}_{\mu\nu} X^\mu\P*X^\nu
}
in the boundary action. Its conformal properties are the same as for the field strength $F$. Since the associated vertex operator has conformal weight 1, ${\cal G}$ is a candidate for a modulus of the theory. In fact the radius deformations (\ref{raddef}) are examples for this.

Another interesting example, which das been discussed in \cite{Gerasimov:2001pg}, is the case where the string field at the boundary acquires a constant expectation value $X^\mu=a^\mu$. The terms in the action are of the form $a^\mu d *X_\mu$. In this case the corresponding operators have an interpretation as displacement operators generating shifts in the positions of D-branes.

In terms of the string $\sigma$-model, the transformation $X\to*X$ has a well known interpretation as a worldsheet implementation of T-duality, at least for flat target spaces. Motivated by this, the so-called doubled formalism has been developed, which has been used to describe string theory on non-geometric backgrounds \cite{Dabholkar:2005ve,Hull:2006va}. Also, arguments coming from cubic string field theory have been provided that motivate the introduction of dual fields \cite{Hull:2009mi}. Another motivation comes from the idea to declare T-duality to be a fundamental property of string theory. On the level of the $\sigma$-model, this can be achieved manually by trying to find a T-invariant action which involves $X$ and $*X$ \cite{Tseytlin:1990nb}. 

In our framework dual fields arise very naturally and should in fact be included as boundary terms in BSFT.
Since factorization properties of the path-integral allow focussing on the worldsheet boundary only, the situation here is different from the usual approaches to doubled field theory, where the bulk plays an important role. Still, the similarities are striking, and it will be interesting to understand the connections between the two approaches.

\section*{Acknowledgements}

We thank the organizers of the SFT09 conference in Moscow, as well as the Steklov Mathematical Institute for hospitality.
This work has been supported by an EURYI grant.

\end{document}